\documentclass[altaffilletter,aps,nofootinbib,twocolumn,prd,eqsecnum,showpacs,showkeys,preprintnumbers,superscriptaddress]{revtex4}
\usepackage[caption=false]{subfig}
\usepackage{graphicx}
\usepackage{amsmath}
\usepackage{amsfonts}
\usepackage{amssymb}
\usepackage{mathtools}
\usepackage{color}
\usepackage{bm}
\usepackage{mathrsfs}
\usepackage{epstopdf}
\usepackage{url}
\usepackage{footnote}
\usepackage{textcomp}
\makeatletter
\newcommand*{\rom}[1]{\expandafter\@slowromancap\romannumeral #1@}
\makeatother

\begin{document}

\title{Black hole solutions in mimetic Born-Infeld gravity}
\author{Che-Yu Chen}
\email{b97202056@gmail.com}
\affiliation{Department of Physics and Center for Theoretical Sciences, National Taiwan University, Taipei, Taiwan 10617}
\affiliation{LeCosPA, National Taiwan University, Taipei, Taiwan 10617}

\author{Mariam Bouhmadi-L\'{o}pez}
\email{mariam.bouhmadi@ehu.eus}
\affiliation{Department of Theoretical Physics, University of the Basque Country UPV/EHU, P.O.~Box 644, 48080 Bilbao, Spain}
\affiliation{IKERBASQUE, Basque Foundation for Science, 48011 Bilbao, Spain}

\author{Pisin Chen}
\email{pisinchen@phys.ntu.edu.tw}
\affiliation{Department of Physics and Center for Theoretical Sciences, National Taiwan University, Taipei, Taiwan 10617}
\affiliation{LeCosPA, National Taiwan University, Taipei, Taiwan 10617}
\affiliation{Kavli Institute for Particle Astrophysics and Cosmology, SLAC National Accelerator Laboratory, Stanford University, Stanford, CA 94305, USA}

\date{\today}

\begin{abstract}
The vacuum, static, and spherically symmetric solutions in the mimetic Born-Infeld gravity are studied. The mimetic Born-Infeld gravity is a reformulation of the Eddington-inspired-Born-Infeld (EiBI) model under the mimetic approach. Due to the mimetic field, the theory contains non-trivial vacuum solutions different from those in Einstein gravity. We find that with the existence of the mimetic field, the spacelike singularity inside a Schwarzschild black hole could be altered to a lightlike singularity, even though the curvature invariants still diverge at the singularity. Furthermore, in this case, the maximal proper time for a timelike radially-infalling observer to reach the singularity is found to be infinite.

\end{abstract}


\maketitle

\section{Introduction}
One of the most fascinating characteristics of Einstein General Relativity (GR) is that GR permits the existence of black holes \cite{caroll,blackhole,hawkingellis}. It is extremely surprising that from such a complicated gravitational theory with highly non-linear and coupled differential equations, one can extract exact solutions by merely assuming some simple while physically reasonable assumptions. For instance, a Schwarzschild black hole stands for an exact solution to GR in a vacuum, static and spherically symmetric configuration. Furthermore, according to our current knowledge of astrophysics, at the later stage of the evolution of a stellar object which is dense and massive enough, nothing can stop the gravitational collapse of the object and it would inevitably end up in a black hole \cite{Adler:2005vn,Chen:2017pkl}. Besides, it is nowadays commonly accepted that there is a supermassive black hole in the center of any galaxy (including our Milky Way), even though the origin of this kind of black holes still lacks proper theoretical explanations.

However, GR not only predicts the existence of black holes, but also forecasts the existence of a singularity inside them \cite{Hawking:1969sw}, even if the singularity is hidden inside an event horizon. At the singularity, the curvature of spacetime diverges and all the geodesics are unable to be extended beyond that. According to GR, an infalling observer would take a finite proper time to cross the event horizon, and beyond that the observer would continue to fall until he reaches the singularity. In fact, this \textit{unfortunate} observer would be even \textit{spaghettified} before reaching the singularity due to the strong tidal forces acting upon him \cite{blackhole}. 

The existence of singularities usually implies the incompleteness of the underlying theory. Given that GR is a purely classical theory, it is expected that some quantum effects or a self-consistent quantum theory of gravity are needed near these classical singularities, and that these singularities may be ameliorated if quantum gravity effects are considered. However, so far a complete and self-consistent quantum theory of gravity remains elusive. We therefore follow a different approach in this work. We will consider an extended theory of gravity, which can be viewed as an effective theory of a full quantum theory of gravity, and expect that, at least at the classical level, the black hole singularity predicted by GR would be altered in this extended theory of gravity \cite{Capozziello:2011et}. 

A particularly interesting attempt following this line is the proposal of the Eddington-inspired-Born-Infeld (EiBI) theory \cite{Banados:2010ix}. The theory contains a Born-Infeld structure in the gravitational action and is able to cure the big bang singularity in the early universe \cite{Banados:2010ix,Scargill:2012kg}. Furthermore, the EiBI theory reduces to GR in vacuum but deviates from it in the presence of matter. For the spherically symmetric configuration, the integral form of the metric function of an electrically charged black hole was firstly given in Ref.~\cite{Banados:2010ix}. Afterwards, the exact expression of the metric function was derived in Refs.~\cite{Sotani:2014lua,Wei:2014dka,Sotani:2015ewa}, where some general properties and the strong gravitational lensing of such black holes were also studied. Besides, some electrically charged solutions for a negative Born-Infeld coupling constant could be interpreted as a wormhole solution \cite{Olmo:2013gqa} and the solutions are geodesically complete \cite{Olmo:2015bya,Olmo:2015dba}. The accretion process of the EiBI black hole and its consequences were discussed in Ref.~\cite{Avelino:2015fve}. In addition, the spherically symmetric solutions in the EiBI theory coupled with Born-Infeld electrodynamics were studied in Ref.~\cite{Jana:2015cha}. When considering general anisotropic fluids (the standard Maxwell field can be regarded as a special case of this fluid) coupled with the EiBI theory, some wormhole solutions and nonsingular naked compact objects can be obtained \cite{Harko:2013aya,Menchon:2017qed,Shaikh:2015oha,Tamang:2015tmd}. Finally, a geodesically complete, non-rotating and charged BTZ black hole in $2+1$ dimensions of the EiBI gravity was found in Ref.~\cite{Bazeia:2016rlg} (see also Ref.~\cite{BeltranJimenez:2017doy} for a recent review on Born-Infeld type of gravity).

As mentioned before, the EiBI theory reduces to GR in vacuum, so the singularity in a Schwarzschild and a Kerr black hole is still unavoidable. In this regard, we will shift to consider the mimetic Born-Infeld gravity, which was proposed in our recent paper \cite{Bouhmadi-Lopez:2017lbx}. In the mimetic Born-Infeld gravity, the EiBI action is combined with the mimetic formulation and the theory contains non-trivial vacuum solutions. The mimetic formulation was firstly applied in Ref.~\cite{Chamseddine:2013kea} to Einstein-Hilbert action to construct non-trivial vacuum solutions. Such solutions can mimic the behavior of dark matter in the cosmological level. Some relevant cosmological and astrophysical applications of the mimetic model can be found in Refs.~\cite{Chamseddine:2014vna,Saadi:2014jfa,Matsumoto:2015wja,Ijjas:2016pad,Matsumoto:2016rsa,Myrzakulov:2015sea,Myrzakulov:2015kda,Astashenok:2015qzw,Nojiri:2014zqa,Astashenok:2015haa,Myrzakulov:2015qaa,Golovnev:2013jxa,Chaichian:2014qba,Deruelle:2014zza,Momeni:2014qta,Arroja:2015wpa,Rabochaya:2015haa,Myrzakulov:2016hrx,Arroja:2015yvd,Cognola:2016gjy,Vagnozzi:2017ilo} (see also Ref.~\cite{Sebastiani:2016ras} for a nice review on the mimetic model).

Essentially, in the mimetic Born-Infeld gravity, the mimetic formulation generates a new branch of vacuum solutions, and these solutions could be somewhat smoothen due to the Born-Infeld structure in the gravitational sector. In Ref.~\cite{Bouhmadi-Lopez:2017lbx}, we have shown that this theory could, to some extent, remove the initial big bang singularity and provide several nonsingular primordial cosmological solutions in a vacuum universe. Therefore, it would be interesting to investigate the vacuum, static and spherically symmetric solutions in the mimetic Born-Infeld model and to study how the interior structure, especially the spacelike singularity, of a Schwarzschild black hole could be modified\footnote{Our work should be distinguished from that in recent papers \cite{Chamseddine:2016uef,Chamseddine:2016ktu} where the big bang singularity \cite{Chamseddine:2016uef} and the Schwarzschild singularity \cite{Chamseddine:2016ktu} are shown to be avoidable by considering a different gravitational theory. In these two papers, the authors combined the mimetic formulation with the standard Einstein-Hilbert action, while a Born-Infeld type function of the d'Alembertian of the mimetic scalar field was introduced. This is the so-called limiting curvature idea in GR.}.

This paper is outlined as follows. In section~\ref{sectII} we briefly review the mimetic Born-Infeld model proposed in Ref.~\cite{Bouhmadi-Lopez:2017lbx}, including the action and the equations of motion. In section~\ref{secIII}, we study the vacuum, static and spherically symmetric solution in this theory. More attention is paid to the behaviors of the interior geometry. The radially infalling proper time of a timelike observer to reach the singularity, and the causal structure of the solution are investigated. Finally, we present our conclusions in section~\ref{conclu}.

\section{Equations of motion}\label{sectII}
The mimetic formulation was proposed for the first time in the seminal paper \cite{Chamseddine:2013kea}, within the context of GR, to explain the mysterious dark matter component at the cosmological level. This formulation is based on a redefinition of the physical metric $g_{\mu\nu}$ such that \cite{Chamseddine:2013kea}:
\begin{equation}
g_{\mu\nu}=-(\tilde{g}^{\alpha\beta}\partial_\alpha\phi\partial_\beta\phi)\tilde{g}_{\mu\nu},
\label{gg}
\end{equation}
where $\tilde{g}_{\mu\nu}$ and $\phi$ are the conformal auxiliary metric and the mimetic scalar field, respectively. On the above equation, $\tilde{g}^{\mu\nu}$ is the inverse of $\tilde{g}_{\mu\nu}$. The parametrization \eqref{gg} respects the conformal invariance of the theory in the sense that the theory is invariant under the conformal transformation $\tilde{g}_{\mu\nu}\rightarrow \Omega^2(x_\alpha)\tilde{g}_{\mu\nu}$, where $\Omega(x_{\alpha})$ is an arbitrary function of the spacetime coordinates. 

Instead of the Einstein-Hilbert action applied in Ref.~\cite{Chamseddine:2013kea}, the mimetic Born-Infeld gravity, which was proposed in our recent work \cite{Bouhmadi-Lopez:2017lbx}, is based on the EiBI action and on the mimetic formulation:
\begin{equation}
\mathcal{S}_{EiBI}=\frac{2}{\kappa}\int d^4x\Big[\sqrt{|g_{\mu\nu}+\kappa R_{\mu\nu}(\Gamma)|}-\lambda\sqrt{-g}\Big]+\mathcal{S}_m(g,\psi),
\label{startaction}
\end{equation}
where $\mathcal{S}_m$ is the matter Lagrangian coupled only with the physical metric $g_{\mu\nu}$. According to the mimetic formulation, the physical metric $g_{\mu\nu}$ in the action should be written as $g_{\mu\nu}(\phi,\tilde{g}_{\alpha\beta})$ on the basis of the parametrization \eqref{gg}. It should be stressed that for the sake of simplicity, the whole calculations throughout this paper are done in absence of any non-trivial potential $V(\phi)$. Furthermore, the dimensionless constant $\lambda$ quantifies the effective cosmological constant at the low curvature limit. On the other hand, $|g_{\mu\nu}+\kappa R_{\mu\nu}(\Gamma)|$ stands for the absolute value of the determinant of the rank two tensor $g_{\mu\nu}+\kappa R_{\mu\nu}(\Gamma)$, where $R_{\mu\nu}(\Gamma)$ is the symmetric part of the Ricci tensor constructed by the affine connection $\Gamma$. The affine connection is further assumed to be symmetric (torsionless) and independent of the metric $g_{\mu\nu}$ (non-metricity). Finally, $\kappa$ characterizes the theory and has inverse dimensions to that of the cosmological constant. Even though the action of the theory looks seemingly similar to that of the original EiBI theory, the equations of motion as well as their applications could be drastically different because of the presence of the mimetic field, as will be shown later.

In the mimetic Born-Infeld theory, it is the auxiliary metric $\tilde{g}_{\mu\nu}$, the mimetic scalar field $\phi$, and the affine connection $\Gamma$ that should be treated as independent variables. After varying the action, the field equations of $\tilde{g}_{\mu\nu}$, $\phi$ and $\Gamma$ can be written as follows \cite{Bouhmadi-Lopez:2017lbx}
\begin{align}
\mathcal{F}^{\mu\nu}+\mathcal{F}g^{\kappa\mu}g^{\lambda\nu}\partial_\kappa\phi\partial_\lambda\phi&=0,\label{gt}\\
\nabla^g_\kappa(\mathcal{F}\partial^\kappa\phi)=\frac{1}{\sqrt{-g}}\partial_\kappa(\sqrt{-g}\mathcal{F}\partial^\kappa\phi)&=0,\label{phi}\\
\nabla^\Gamma_{\alpha}(g_{\mu\nu}+\kappa R_{\mu\nu})&=0,\label{gamma}
\end{align}
respectively. On the above equations, $\nabla^g_\kappa$ and $\nabla^\Gamma_\alpha$ denote the covariant derivative defined by the metric $g_{\mu\nu}$ and by the affine connection $\Gamma$, respectively. The tensor $\mathcal{F}^{\mu\nu}$ is defined as
\begin{equation}
\mathcal{F}^{\mu\nu}\equiv\frac{\sqrt{|\hat{g}+\kappa \hat{R}|}}{\sqrt{-g}}[(\hat{g}+\kappa \hat{R})^{-1}]^{\mu\nu}-\lambda g^{\mu\nu}+\kappa T^{\mu\nu},
\end{equation}
where $T_{\mu\nu}$ is the energy momentum tensor, and $\mathcal{F}\equiv g_{\mu\nu}\mathcal{F}^{\mu\nu}$. The hat symbolizes a matrix quantity. Eq.~\eqref{gamma} implies that there exists a second auxiliary metric $q_{\mu\nu}\equiv g_{\mu\nu}+\kappa R_{\mu\nu}$ such that $q_{\mu\nu}$ is compatible with the affine connection $\Gamma$. In the original EiBI theory within the Palatini variational principle, there is no mimetic scalar field so the equation of motion of the physical metric $g_{\mu\nu}$ is simply $\mathcal{F}^{\mu\nu}=0$. Therefore, in the mimetic Born-Infeld model, the second term in Eq.~\eqref{gt}, which is a contribution of the mimetic scalar field, results in solutions which are absent in the original EiBI theory. Note that the mimetic scalar field is confined to satisfy the constraint:
\begin{equation}
g^{\mu\nu}\partial_\mu\phi\partial_\nu\phi=-1.
\label{constraintphi}
\end{equation}
This constraint can be derived straightforwardly from the parametrization \eqref{gg}.

To implement the equations of motion, it is more convenient to define a matrix as follows \cite{Olmo:2013gqa}:
\begin{equation}
\hat{\Omega}\equiv\hat{g}^{-1}\hat{q}\,,\qquad \hat{\Omega}^{-1}\equiv\hat{q}^{-1}\hat{g}\,,
\end{equation}
such that $\hat{q}=\hat{g}\hat{\Omega}$. The field equation \eqref{gt} can be written as
\begin{equation}
\sqrt{|\hat{\Omega}|}\hat{\Omega}^{-1}-\lambda\hat{I}+\kappa\hat{T}+\mathcal{F}\hat{K}=0,
\label{matrixeq1}
\end{equation}
where $\hat{T}\equiv T^{\mu\alpha}g_{\alpha\nu}$, $\hat{I}$ is the four-dimensional identity matrix, and $\hat{K}\equiv\partial^{\mu}\phi\partial_{\nu}\phi$. According to the constraint \eqref{constraintphi} it can be seen that the trace of $\hat{K}$ is $\textrm{Tr}(\hat{K})=-1$. Additionally, the field equation \eqref{gamma} can be written as
\begin{equation}
{R^\mu}_\nu[q]\equiv \hat{q}^{-1}\hat{R}=\frac{1}{\kappa}(\hat{I}-\hat{\Omega}^{-1}).
\label{matrixeq2}
\end{equation}

Before closing this section, we would like to stress that the field equations \eqref{gt}, \eqref{phi} and \eqref{gamma} can be obtained by varying an alternative action
\begin{align}
\mathcal{S}_{a}=&\,\frac{1}{2}\int d^4x\sqrt{-q}\Big[R[q]-\frac{2}{\kappa}+\frac{1}{\kappa}\Big(q^{\alpha\beta}g_{\alpha\beta}-2\sqrt{\frac{g}{q}}\lambda\Big)\Big]\nonumber\\&+S_m(g,\psi),
\label{alternativeaction}
\end{align}
within the mimetic setup with respect to $\tilde{g}_{\mu\nu}$, $\phi$ and $q_{\mu\nu}$. This fact further confirms the equivalence of this action and action \eqref{startaction}. In the original EiBI theory, this alternative action was firstly discovered in Ref.~\cite{Delsate:2012ky} and then applied in Refs.~\cite{Bouhmadi-Lopez:2016dcf,Arroja:2016ffm,Albarran:2017swy} in the context of quantum cosmology. The equivalence between these two actions is still valid in the mimetic setup and we will explore it in the context of quantum cosmology in a forthcoming paper Ref.~\cite{ongoingworksss}.

\section{Spherically symmetric solution}\label{secIII}
It is a well known fact that the EiBI theory is equivalent to GR in vacuum, hence the theory shares the same vacuum solution of GR. However, according to the equations of motion \eqref{gt}, \eqref{phi} and \eqref{gamma}, the mimetic Born-Infeld theory contains a non-trivial vacuum solution, which is absent in GR, because of the presence of the mimetic field. In our accompanying paper \cite{Bouhmadi-Lopez:2017lbx}, we have proven that this model could, to some extent, remove the initial big bang singularity and provide several primordial cosmological solutions in absence of matter. Therefore, it would be interesting to investigate the vacuum, static, and spherically symmetric solutions in the mimetic Born-Infeld model and to study how the Schwarzschild solution could be altered, especially the spacelike \textit{center} of a Schwarzschild black hole, by the existence of the mimetic field.

We consider a vacuum spacetime in which $T_{\mu\nu}=0$ and assume a static and spherically symmetric ansatz:
\begin{equation}
ds^2=-\psi^2(r)f(r)dt^2+\frac{1}{f(r)}dr^2+r^2d\Omega^2,
\label{metricg}
\end{equation}
where $d\Omega^2=d\vartheta^2+\sin^2{\vartheta}d\varphi^2$. The mimetic scalar field $\phi$ depends only on $r$ within this configuration. Therefore, the constraint \eqref{constraintphi} can be written as
\begin{equation}
\Big(\frac{d\phi}{dr}\Big)^2=-\frac{1}{f(r)},
\label{hh}
\end{equation}
and the mimetic scalar field is an imaginary (real) field if $f(r)$ is positive (negative). Furthermore, the matrix $\hat{K}=\partial^{\mu}\phi\partial_{\nu}\phi$ is
\begin{equation}
\hat{K}=
\begin{bmatrix}
    0       & 0 & 0 & 0 \\
    0       & -1 & 0 & 0 \\
   0       & 0 & 0 & 0 \\
    0       & 0 & 0 & 0 
\end{bmatrix}.
\end{equation}
From Eq.~\eqref{matrixeq1}, we obtain
\begin{equation}
\hat{\Omega}=\lambda
\begin{bmatrix}
    X(r)       & 0 & 0 & 0 \\
    0       & \frac{1}{X(r)} & 0 & 0 \\
   0       & 0 & X(r) & 0 \\
    0       & 0 & 0 & X(r) 
\end{bmatrix},
\end{equation}
where the function $X(r)$ is defined as
\begin{equation}
X(r)\equiv\sqrt{1+\frac{\mathcal{F}(r)}{\lambda}}.
\end{equation}
In absence of the mimetic field, we have $\mathcal{F}(r)=0$ and $X(r)=1$ for all $r$. The more the value of $X(r)$ deviates from unity, the more the mimetic field contributes to the dynamics of the system. Therefore, the function $X(r)$ can essentially be interpreted as a measure of the impact of the mimetic field in the theory.  

According to the map $\hat{q}=\hat{g}\hat{\Omega}$, the second auxiliary metric $q_{\mu\nu}$, which is compatible with the affine connection, reads
\begin{equation}
ds_q^2=-\lambda \psi^2(r)f(r)X(r)dt^2+\frac{\lambda}{X(r)f(r)}dr^2+\lambda r^2X(r)d\Omega^2.
\label{qoriginal}
\end{equation}
To proceed, we choose a different coordinate system in which the auxiliary metric can be written as
\begin{equation}
ds_q^2=-G^2(x)H(x)dt^2+\frac{1}{H(x)}dx^2+x^2d\Omega^2.
\label{qnew}
\end{equation}
Comparing the expressions \eqref{qoriginal} and \eqref{qnew}, we have the following identities
\begin{equation}
G^2(x)H(x)=\lambda\psi^2(r)f(r)X(r)\,,\qquad \Big(\frac{dx}{dr}\Big)^2=\frac{\lambda H(x)}{X(r)f(r)}\,,\label{identity}
\end{equation}
and
\begin{equation}
x^2=\lambda r^2X(r)\,.\label{identity2}
\end{equation}

Considering the non-vanishing components of Eq.~\eqref{matrixeq2} and writing them in terms of $x$, we obtain
\begin{align}
&\,H\Big[\frac{1}{x}\Big(\frac{2G'}{G}+\frac{H'}{H}\Big)+\frac{3}{2}\frac{G'H'}{GH}+\frac{G''}{G}+\frac{1}{2}\frac{H''}{H}\Big]\nonumber\\
=&\,\frac{1}{\kappa}\Big(\frac{1}{\lambda X}-1\Big),\label{1}\\
&\,H\Big(-\frac{1}{x}\frac{H'}{H}-\frac{3}{2}\frac{G'H'}{GH}-\frac{G''}{G}-\frac{1}{2}\frac{H''}{H}\Big)=\frac{1}{\kappa}\Big(1-\frac{X}{\lambda}\Big),\label{2}\\
&\,\frac{1}{x^2}-H\Big[\frac{1}{x^2}+\frac{1}{x}\Big(\frac{G'}{G}+\frac{H'}{H}\Big)\Big]=\frac{1}{\kappa}\Big(1-\frac{1}{\lambda X}\Big)\label{3},
\end{align}
where the prime denotes the derivative with respect to $x$. Note that even though $X$ is initially introduced as a function of $r$, it can be expressed as a function of $x$ because $x$ is intrinsically a function of $r$ through Eqs.~\eqref{identity} and \eqref{identity2}, and vice versa. 

After some calculations, we obtain
\begin{equation}
\frac{d}{dx}(xH(x))=1-\frac{x^2}{\kappa}+\frac{1}{2\kappa\lambda}\Big(\frac{1}{X}+X\Big)x^2.
\end{equation}
This equation can be rewritten as follows
\begin{equation}
H(x)=1-\frac{1}{3\kappa}x^2+\frac{c_1}{x}+\frac{\xi(x)}{x},
\label{73}
\end{equation}
where
\begin{equation}
\xi(x)\equiv \frac{1}{2\kappa\lambda}\int\Big(\frac{1}{X}+X\Big)x^2dx,
\end{equation}
and $c_1$ is an integration constant. On the other hand, Eq.~\eqref{3} can be written as
\begin{equation}
\frac{G'}{G}=\frac{1}{xH}\Big[1-\frac{x^2}{\kappa}+\frac{x^2}{\kappa\lambda X}-\frac{d}{dx}(xH)\Big].
\end{equation}
This equation leads to
\begin{equation}
G^2(x)=c_2\,\textrm{exp}\Bigg[\int\frac{\frac{x^2}{\kappa\lambda}\Big(\frac{1}{X}-X\Big)}{xH(x)}dx\Bigg],
\label{76}
\end{equation}
where $c_2$ is another integration constant. In absence of the mimetic field, i.e., $X=1$, we recover the Schwarzschild-de Sitter solution by choosing $c_1=-\sqrt{\lambda}r_s$, where $r_s$ is the Schwarzschild radius, and $c_2=\lambda$.

To derive the solutions in the presence of the mimetic field, we use the fact that, in addition to Eqs.~\eqref{73} and \eqref{76}, one can obtain from Eq.~\eqref{1} a separate equation governing the behavior of $X$:
\begin{equation}
\frac{3}{x}+\frac{X'(X^2+3)}{X(X^2-1)}+\frac{1-\frac{x^2}{\kappa}+\frac{x^2}{2\kappa\lambda}\Big(\frac{3}{X}-X\Big)}{xH(x)}=0.\label{complicated}
\end{equation}
This equation is trivially satisfied when $X=1$, i.e., in absence of the mimetic field. Note that this equation can be derived by combining Eqs.~\eqref{phi} and \eqref{76} as well.

\subsection{The interior structure}
As mentioned previously, the solutions in the static, vacuum and spherically symmetric geometry reduce to the Schwarzschild-de Sitter solution in absence of the mimetic field, i.e., $X=1$. In this subsection, we will study how the interior geometry of a black hole, especially the singularity, is modified by the presence of the mimetic field. Given that the differential equation \eqref{complicated} is too complicated to be solved analytically, we will resort to numerical methods.

We firstly assume $\lambda=1$, i.e., a vanishing cosmological constant, for the sake of simplicity. After deriving the solutions, we will compare the results with the standard Schwarzschild solution. On a certain radius ($x=x_i$) inside the event horizon, we assume that there is a small amount of the mimetic field and the solutions deviate from the Schwarzschild geometry within this radius, that is, $X\neq 1$ when $x\le x_i$. This particular radius $x_i$ is the point where the initial conditions are imposed. More precisely, we assume $X(x_i)=1+\delta$, and $|\delta|$ can be made rather small. Under this assumption, it can be seen from Eq.~\eqref{complicated} that $X'(x_i)$ is also of the order of $\delta$ and so are its higher derivatives at $x=x_i$. The major goal of this work is to study how a small deviation $\delta$ in the mimetic field would alleviate the spacelike singularity in the interior of a Schwarzschild black hole. In the rest of this work, we will assume a positive $\kappa$ because of the instability problems ubiquitous to a negative $\kappa$ \cite{Avelino:2012ge}.

\begin{figure}[t]
\centering
\graphicspath{{fig/}}
\includegraphics[scale=0.9]{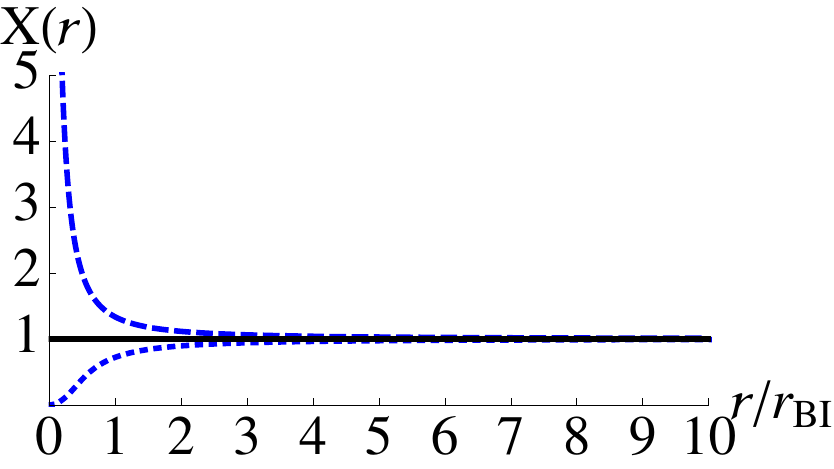}
\caption{$X(r)$ is shown as a function of $r/r_{\textrm{BI}}$, where $r_{\textrm{BI}}\equiv\sqrt{\kappa}$. The dashed curve corresponds to an initial condition $\delta=0.01$, i.e., $\delta>0$, at $x_i=10$. The dotted curve, on the other hand, corresponds to an initial condition $\delta=-0.01$, i.e., $\delta<0$. The solution without the mimetic field, i.e., $X=1$ is shown by the solid line.} 
\label{test123}
\end{figure}

From now on, we normalize the radius by assuming $x\rightarrow x/\sqrt{\kappa}$, and numerically solve Eq.~\eqref{complicated}. Under this normalization, $x$ becomes dimensionless and it can be converted back to the radius $r$ by using the identities \eqref{identity} and \eqref{identity2}. The numerical results of the function $X(r)$ are shown in FIG.~\ref{test123}. The dashed curve is derived by assuming an initial condition $\delta=0.01$ at $x_i=10$, and the dotted curve corresponds to an initial condition $\delta=-0.01$ at $x_i=10$. Note that the qualitative behaviors of the solutions do not depend on the quantitative values of these conditions once the sign of $\delta$ is fixed. The solution in absence of the mimetic field, i.e., the Schwarzschild solution, is simply $X(r)=1$ and it is shown by the solid line. It can be seen that if $X\neq 1$, the solutions deviate from the Schwarzschild solution when $r\rightarrow 0$ and the behaviors of the solutions depend on the sign of $\delta$ chosen at $x_i$. When $r\rightarrow 0$, the approximated behaviors of $X(r)$ can be obtained as follows
\begin{equation}
\begin{dcases}
X(x)\approx b_1x^{-3}\, \\
X(x)\approx b_2x\, 
\end{dcases}\rightarrow
\begin{dcases}
X(r)\approx b_1^{2/5}r^{-6/5}\,, & \mbox{(dashed)} \\
X(r)\approx b_2^2r^2\,,& \mbox{(dotted)} 
\end{dcases}
\label{Xbehavior}
\end{equation}
where $b_1$ and $b_2$ are positive integration constants related to the initial conditions.

\begin{figure}[t]
\centering
\graphicspath{{fig/}}
\includegraphics[scale=0.9]{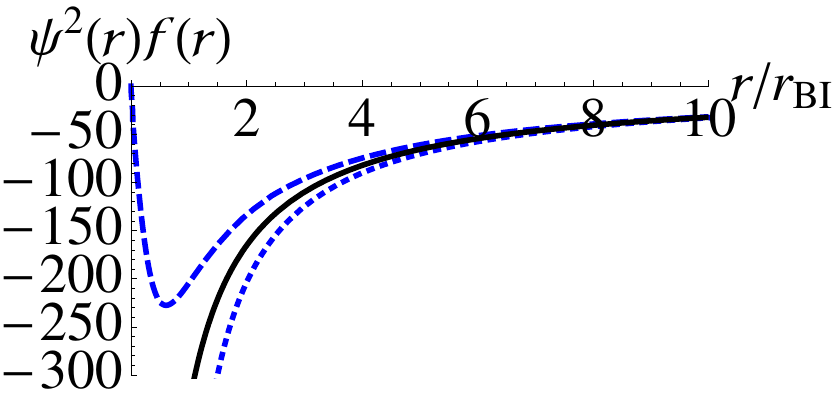}
\includegraphics[scale=0.9]{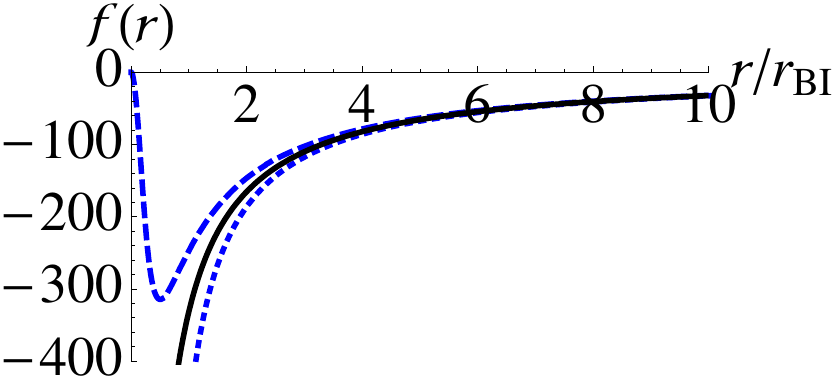}
\caption{The numerical results of the metric functions $\psi^2(r)f(r)$ (upper) and $f(r)$ (lower) are shown as functions of $r/r_\textrm{BI}$. The standard Schwarzschild solution is shown by the solid curves.} 
\label{test211}
\end{figure}

Furthermore, the metric functions can be obtained by numerically calculating Eqs.~\eqref{identity}, with the numerical results of $X(r)$. The results are shown in FIG.~\ref{test211}, where the functions $\psi^2(r)f(r)$ (top) and $f(r)$ (bottom) are shown as functions of $r$. The standard Schwarzschild solution is also shown by the solid curve. According to FIG.~\ref{test211}, it can be seen again that the solutions deviate significantly from the Schwarzschild solution when $r\rightarrow 0$. The approximated solutions when $r\rightarrow 0$ for the dashed and dotted curves can be obtained as follows:
\begin{equation}
\begin{dcases}
\psi^2(r)f(r)\approx \frac{5}{4}\frac{b_1^{2/5}}{\kappa}\frac{r^{4/5}}{\ln{r}}\,,\qquad f(r)\approx\frac{5}{4}\frac{r^2}{\kappa}\ln{r}, & \mbox{(dashed)} \\
\psi^2(r)f(r)\approx-\frac{r_s}{b_2^3}r^{-4}\,,\qquad f(r)\approx-\frac{r_s}{4b_2^5}r^{-6}\,. & \mbox{(dotted)} 
\end{dcases}
\label{singularg}
\end{equation}

In addition, it can be shown that the Ricci scalar $R[g]\equiv g^{\mu\nu}R_{\mu\nu}[g]$ and Kretschmann invariant $K[g]\equiv R_{abcd}[g]R^{abcd}[g]$ constructed from the physical metric $g_{\mu\nu}$, whose approximated behaviors are given in Eqs.~\eqref{singularg}, diverge at $r\rightarrow 0$:
\begin{equation}
\begin{dcases}
R[g]\approx \frac{2}{r^2}\,,\qquad K[g]\approx \frac{4}{r^4}, & \mbox{(dashed)} \\
R[g]\approx 6\Big(\frac{r_s}{b_2^5}\Big)r^{-8}\,,\qquad K[g]\approx 684\Big(\frac{r_s}{b_2^5}\Big)^2r^{-16}. & \mbox{(dotted)} 
\end{dcases}
\label{curvature2}
\end{equation}
Therefore, there is a curvature singularity at $r=0$ for the two choices of initial conditions. 

\subsection{A radially infalling observer}
According to the numerical and approximated solutions shown in the previous subsection, the vacuum, static, and spherically symmetric geometry differs from the Schwarzschild black hole in the mimetic Born-Infeld model. Although the existence of a curvature singularity at $r\rightarrow 0$ seems unavoidable, the behaviors of the metric functions could differ significantly. An interesting and important quantity which can be compared with that in the Schwarzschild black hole is the infalling proper time of a timelike observer elapsed to reach the singularity. It is well known that the proper time for a radially infalling observer to reach the Schwarzschild singularity is finite. How this quantity is altered in the mimetic Born-Infeld model deserves some scrutinies. This issue will be addressed in this subsection.

\subsubsection{Killing vectors and constants of motion}
For a spacetime with a given symmetry, there exists a vector $k^{\mu}$ that characterizes the symmetry and satisfies the Killing's equation \cite{blackhole,hawkingellis}
\begin{equation}
\nabla^g_\mu k_{\nu}+\nabla^g_\nu k_{\mu}=0.
\label{Killingeq}
\end{equation}
A vector field $k^{\mu}$ satisfying this equation is called a Killing vector. 

Let us consider a geodesic curve $x^{\mu}=x^{\mu}(\tau)$ in the spacetime described by\footnote{In a metric-affine theory like the one we are considering, one can as well define a geodesic curve described by the affine connection. However, given that in our theory the matter sector does not couple to the affine connection, test particles should follow the geodesics defined by $g_{\mu\nu}$ and that is the reason why we choose these geodesics here. Note that within this framework, the Einstein equivalence principle is satisfied \cite{Capozziello:2011et}.} $g$ and define $u^\mu=dx^{\mu}/d\tau$ to be the tangent vector to the curve. Because the curve is a geodesic, we have $u^{\nu}\nabla^g_{\nu}u^{\mu}=0$. The rate of change of the quantity $u^{\mu}k_{\mu}$ along the geodesic curve is
\begin{align}
&\frac{d}{d\tau}(u^{\mu}k_{\mu})\nonumber\\
=&\,u^{\nu}\nabla^g_{\nu}(u^{\mu}k_{\mu})\nonumber\\
=&\,(u^{\nu}\nabla^g_{\nu}u^{\mu})k_{\mu}+u^{\nu}u^{\mu}\nabla^g_{\nu}k_{\mu}\nonumber\\
=&\,\frac{1}{2}u^{\nu}u^{\mu}(\nabla^g_{\nu}k_{\mu}+\nabla^g_{\mu}k_{\nu})\nonumber\\
=&\,0.\label{3.23}
\end{align}
We have used the Killing's equation \eqref{Killingeq} and the geodesic equation $u^{\nu}\nabla^g_{\nu}u^{\mu}=0$ to arrive to the result \eqref{3.23}. Therefore, $u^{\mu}k_{\mu}$ is a constant of motion along the geodesic curve and it is intrinsic to the Killing vector $k^{\mu}$ associated with the symmetry in the spacetime.

\subsubsection{Infalling proper time of a timelike observer}
For a static and spherically symmetric metric given in Eq.~\eqref{metricg}, there are two killing vectors: $k^{\mu}=(1\,,0\,,0\,,0)$ and $(0\,,0\,,0\,,1)$. The constants of motion along the geodesic curve are
\begin{align}
\psi^2(r)f(r)\frac{dt}{d\tau}=&\,\varepsilon,\\
r^2\frac{d\varphi}{d\tau}=&\,L,
\end{align}
respectively. On the above equations, $\varepsilon$ and $L$ can be regarded as the conserved energy and the angular momentum of the system. In this regard, the timelike geodesic equation can be derived by using $g_{\mu\nu}u^{\mu}u^{\nu}=-1$ and it reads
\begin{equation}
-\frac{\varepsilon^2}{\psi^2(r)f(r)}+\frac{1}{f(r)}\Big(\frac{dr}{d\tau}\Big)^2+\frac{L^2}{r^2}=-1.
\label{326}
\end{equation}
Note that we have considered the motion on the plane $\vartheta=\pi/2$. For a radial motion, we have $L=0$.

\begin{figure}[t]
\centering
\graphicspath{{fig/}}
\includegraphics[scale=0.9]{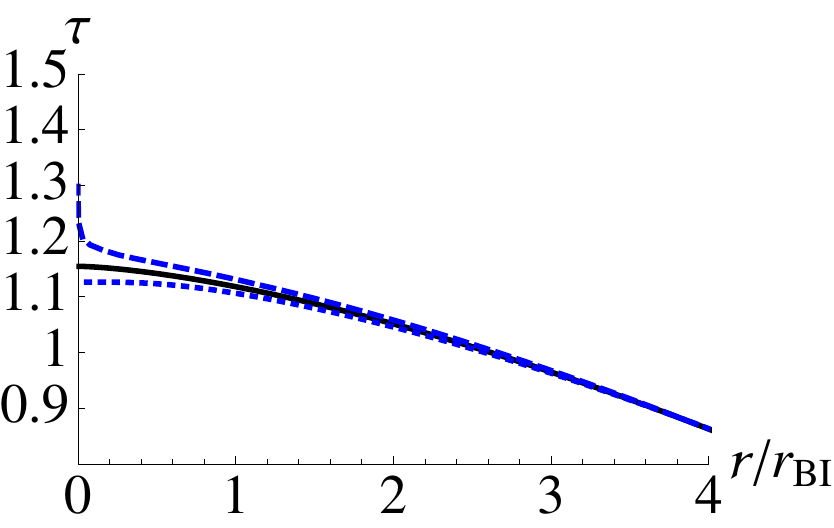}
\caption{The infalling proper time $\tau(r)$ for $\varepsilon=1$ is shown as a function of $r/r_\textrm{BI}$. The dashed, dotted and the solid curves correspond to $\delta=0.01$, $\delta=-0.01$ and $\delta=0$ (GR) at $x_i=10$, respectively.} 
\label{propertime1}
\end{figure}

\begin{figure}[t]
\centering
\graphicspath{{fig/}}
\includegraphics[scale=0.9]{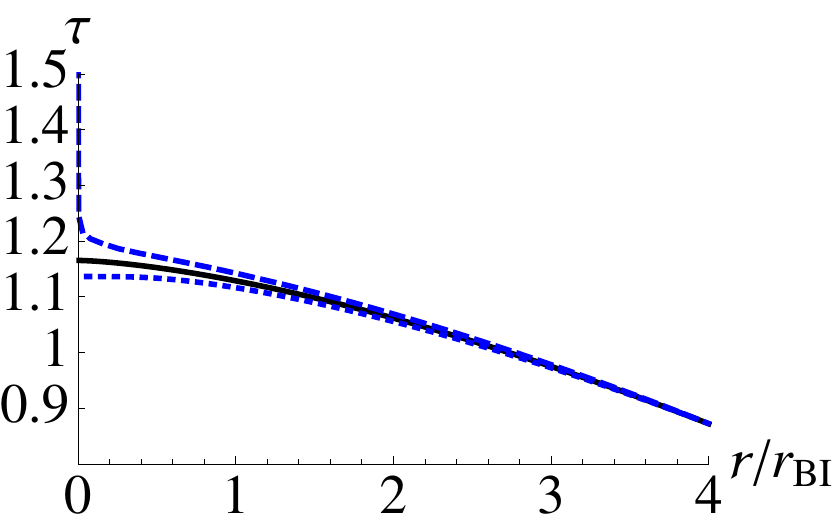}
\caption{The maximal infalling proper time $\tau(r)$ for $\varepsilon=0$ is shown as a function of $r/r_\textrm{BI}$. The dashed, dotted and the solid curves correspond to $\delta=0.01$, $\delta=-0.01$ and $\delta=0$ (GR) at $x_i=10$, respectively.} 
\label{propertime0}
\end{figure}

Then, we consider two different cases to analyze the infalling proper time: (i) $\varepsilon=1$ and (ii) $\varepsilon=0$. The first case, $\varepsilon=1$, corresponds to a situation in which an observer is at rest at infinity and falls freely into the black hole. In the second case, $\varepsilon=0$, the observer is initially at rest on the event horizon. The proper time for the second case is called maximal infalling proper time \cite{caroll}. We use the numerical results of the metric functions in the previous subsections and derive the infalling proper time $\tau(r)$ numerically for these two cases. We assume that the observer starts to count his/her time when crossing $x_i$, that is, $\tau(x_i)=0$. The results of the first case ($\varepsilon=1$) and of the second case ($\varepsilon=0$) are shown in FIG~\ref{propertime1} and FIG~\ref{propertime0}, respectively. One can see that for $\varepsilon=1$, the infalling proper time to reach the singularity is finite for both choices of initial conditions, even though the proper time to reach the singularity is slightly \textit{postponed} for the solution described by the dashed curve ($\delta>0$) (see FIG~\ref{propertime1}), compared with the GR counterpart. On the other hand, we find that, according to the dashed curve ($\delta>0$) in FIG.~\ref{propertime0}, the maximal infalling proper time ($\varepsilon=0$) to reach the singularity is infinite. This can be briefly elucidated as follows 
\begin{align}
\tau(r)_{\varepsilon=0,\,\delta>0}=&\int\frac{-dr}{\sqrt{-f(r)}}\approx\int\frac{-dr}{r\sqrt{-\ln{r}}}\nonumber\\
=&2\sqrt{-\ln{r}}\rightarrow\infty,
\end{align}
when $r\rightarrow0$. 
However, the maximal infalling proper time for the solution described by the dotted curve ($\delta<0$) is even smaller than its GR counterpart (see FIG~\ref{propertime0}). 

For completeness, we discuss what happens to a lighlike observer freely falling toward the singularity. If we consider an observer who follows a lightlike free falling geodesic, the geodesic equation can be obtained by replacing the right hand side of Eq.~\eqref{326} with zero. The equation reads
\begin{equation}
\frac{1}{\varepsilon^2}\Big(\frac{dr}{d\tau}\Big)^2=\frac{1}{\psi^2(r)}-\frac{b^2}{r^2}f(r),
\end{equation}
where $b\equiv L/\varepsilon$ can be interpreted to be an apparent impact parameter as seen from asymptotic infinity. We find that this lighlike observer would take a finite proper time to reach the curvature singularity, irrespective of the value of $b$, $\varepsilon$, and the sign of $\delta$.

\subsection{The causal structure of the singularity at $r=0$}
\begin{figure}[t]
\centering
\graphicspath{{fig/}}
\includegraphics[scale=0.5]{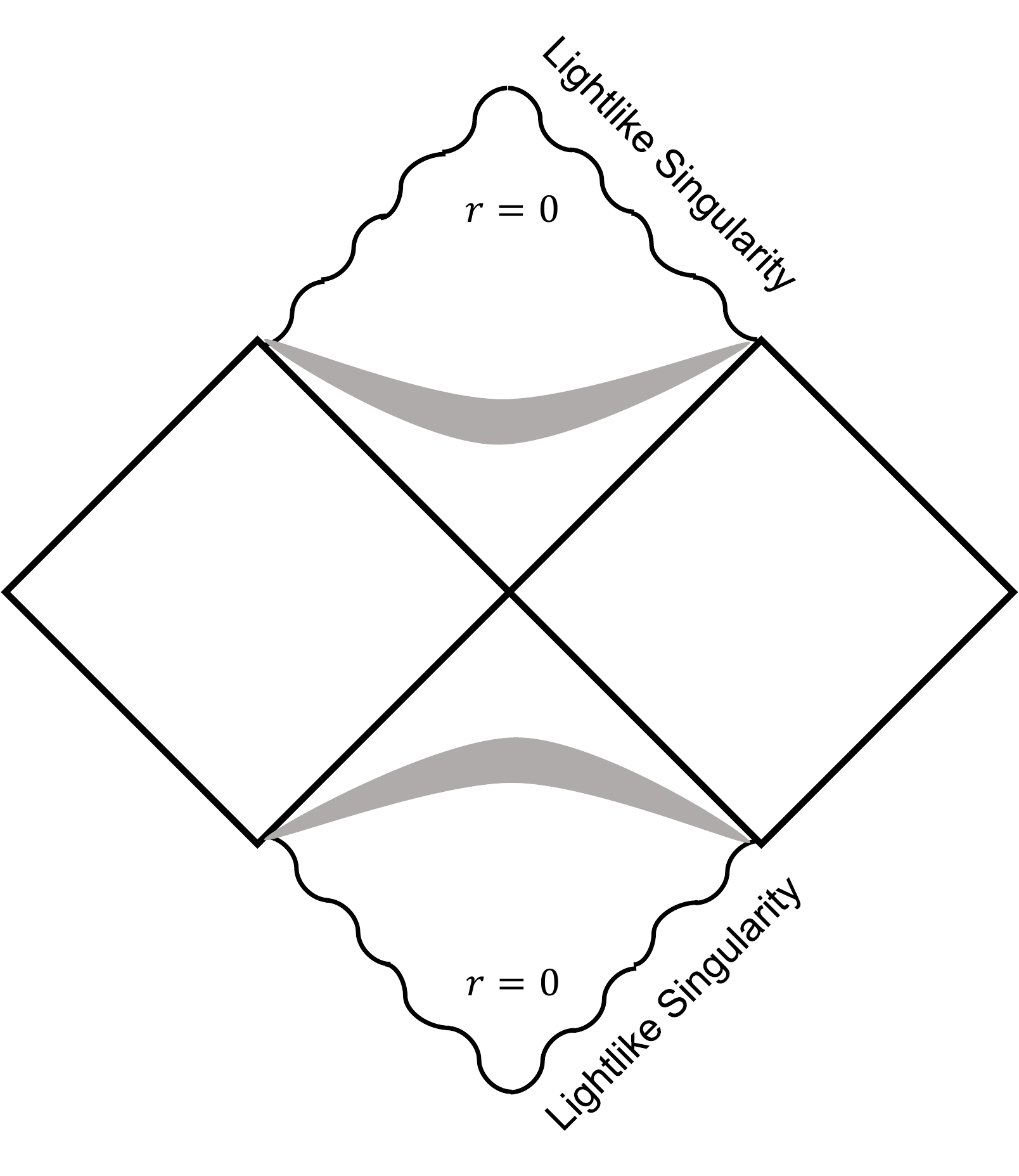}
\caption{The Penrose diagram of the lightlike singularity in the black hole for $\delta>0$. The curly lines indicate the lightlike singularities. The shadowed regions refer to the spacetime patches around $x_i$, if we match a Schwarzschild spacetime and the interior spacetime with non-vanishing mimetic fields.} 
\label{penrose}
\end{figure}

Another important property of the curvature singularity is its causal structure. In this subsection, we will determine the causal structure of the black hole singularity in the mimetic Born-Infeld gravity in more detail. We firstly focus on the $tr$ plane and introduce the following set of new coordinates
\begin{equation}
\bar{t}=t\,,\qquad d\bar{r}=\frac{dr}{\psi(r)f(r)}.
\end{equation}
The metric line element can be written as
\begin{equation}
ds^2=-\psi^2(r)f(r)(d\bar{t}^2-d\bar{r}^2).
\end{equation}
Next, we further define a new coordinate
\begin{equation}
\bar u=e^{\bar{A}(\bar{t}+\bar{r})}\,,\qquad \bar v=-e^{-\bar{A}(\bar{t}-\bar{r})}\,,
\end{equation}
such that
\begin{equation}
d\bar{u}d\bar{v}=\bar{A}^2e^{2\bar{A}\bar{r}}(d\bar{t}^2-d\bar{r}^2),
\end{equation}
where $\bar{A}$ is a constant. Finally, we define a new timelike and a spacelike coordinate as follows: $\bar{T}=(\bar{u}-\bar{v})/2$ and $\bar{X}=(\bar{u}+\bar{v})/2$, such that $-d\bar{T}^2+d\bar{X}^2=d\bar{u}d\bar{v}$. The line element becomes
\begin{equation}
ds^2=-\psi^2(r)f(r)e^{-2\bar{A}\bar{r}}\bar{A}^{-2}(-d\bar{T}^2+d\bar{X}^2),
\end{equation}
and we have
\begin{equation}
\bar{T}^2-\bar{X}^2=-\bar{u}\bar{v}=e^{2\bar{A}\bar{r}}.
\end{equation}
For the solutions with initial conditions $\delta>0$ (dashed curves), we have
\begin{equation}
ds^2=-\frac{5}{4}\frac{b_1^{2/5}}{\kappa\bar{A}^2}\frac{r^{4/5}}{\ln{r}}\textrm{exp}\Big(\frac{4\kappa\bar{A}}{b_1^{1/5}}r^{-2/5}\Big)(-d\bar{T}^2+d\bar{X}^2),
\label{conformalg}
\end{equation}
and
\begin{equation}
\bar{T}^2-\bar{X}^2=\textrm{exp}\Big(-\frac{4\kappa\bar{A}}{b_1^{1/5}}r^{-2/5}\Big),
\end{equation}
when $r\rightarrow 0$. To see the behavior of the geometry near $r\rightarrow0$ more clearly, we have to assume a positive $\bar{A}$ such that the prefactor in the line element \eqref{conformalg} does not vanish near $r\rightarrow 0$. Note that the corresponding $\bar{A}$ in the Schwarzschild spacetime expressed in the Kruskal-Szekeres coordinates is $\bar{A}=1/(2r_s)$. Therefore, if $\delta>0$, we have 
\begin{equation}
\bar{T}^2-\bar{X}^2=0,
\end{equation}
when $r=0$. This means that the curvature singularity is a lightlike singularity. If we connect the two portions of the spacetime: the interior structure described above ($x\le x_i$) and the Schwarzschild spacetime ($x>x_i$), the causal structure of the lightlike singularity and its corresponding Penrose diagram are depicted in FIG.~\ref{penrose}. 

On the other hand, for the solutions with initial condition $\delta<0$ (dotted curves), we have 
\begin{equation}
ds^2=\frac{r_s}{b_2^3\bar{A}^2}r^{-4}\textrm{exp}\Big(-\frac{2}{3}\frac{b_2^4}{r_s}\bar{A}r^6\Big)(-d\bar{T}^2+d\bar{X}^2),
\label{conformalg2}
\end{equation}
and
\begin{equation}
\bar{T}^2-\bar{X}^2=\textrm{exp}\Big(\frac{2}{3}\frac{b_2^4}{r_s}\bar{A}r^6\Big)\rightarrow 1,
\end{equation}
when $r\rightarrow 0$. Therefore, the singularity in this case is a spacelike singularity, similar to the Schwarzschild singularity. 

\section{conclusion}\label{conclu}
The vacuum, static, and spherically symmetric solutions within the mimetic Born-Infeld gravity are studied. The mimetic Born-Infeld gravity \cite{Bouhmadi-Lopez:2017lbx} consists of a reformulation of the EiBI action, combined with the mimetic formulation. This formulation is based on the reparametrization of the physical metric, i.e., Eq.~\eqref{gg}. As a result, the theory contains non-trivial vacuum solutions. We have shown in Ref.~\cite{Bouhmadi-Lopez:2017lbx} that this theory provides interesting and well-defined cosmological solutions describing the primordial era of the universe. It is then natural to study how the interior structure, or the singularity, of a Schwarzschild black hole, which for the EiBI formulation stands for a vacuum solution, could be altered in the mimetic Born-Infeld gravity.

In absence of the mimetic field $(X=1)$, the theory reduces to GR and the solution is simply the Schwarzschild black hole, if the effective cosmological constant is assumed to be zero $(\lambda=1)$. However, if we impose a small amount of the mimetic field on a certain radius inside the event horizon, i.e., $X(x_i)=1+\delta$, our numerical results indicate that the interior structure of a black hole would be different from that of the Schwarzschild geometry. This deviation becomes quite significant near the origin $r\rightarrow 0$ and the behaviors of the solutions depend on the sign of $\delta$ that we impose on $x_i$. We consider a positive Born-Infeld coupling\footnote{This choice of $\kappa$ $(\kappa>0)$ is motivated by the instability usually present in the EiBI theory with a negative $\kappa$ \cite{Avelino:2012ge}.} $(\kappa>0)$ and find that, if $\delta>0$, the metric functions $\psi^2(r)f(r)$ and $f(r)$ approach zero at the origin. This can be seen in Eqs.~\eqref{singularg} and in the dashed curves in FIG.~\ref{test211}. On the other hand, if $\delta<0$, the metric functions diverge at the origin and drop more rapidly than those do near the Schwarzschild singularity. This can be seen in Eqs.~\eqref{singularg} and in the dotted curves in FIG.~\ref{test211}. We show that, in these two cases, the scalar invariants diverge at the origin and this implies the existence of a curvature singularity at $r=0$.

Furthermore, we consider a timelike observer who moves along the geodesic of the spacetime and radially falls into the black hole. Using the metric functions that we have obtained numerically, we calculate the proper time of this observer to reach the curvature singularity at the origin. We find that if the observer is initially at rest at spatial infinity ($\varepsilon=1$), this observer would take a finite proper time to reach the singularity. This can be seen in FIG.~\ref{propertime1}. On the other hand, if we calculate the maximal infalling proper time by assuming the observer to be initially at rest on the event horizon, that is, $\varepsilon=0$, it would take an infinite (finite) proper time to arrive at the singularity if $\delta>0$ ($\delta<0$). 

Next, we analyze the causal structure of the obtained solutions. We find that if $\delta<0$, the curvature singularity at the origin is spacelike and it is stronger than the Schwarzschild singularity in the sense that the curvature invariants diverge more rapidly in this solution. On the other hand, if $\delta>0$, the curvature singularity at the origin becomes a lightlike singularity and we regard the singularity in this case a weaker singularity in the sense that the maximal proper time of a radially infalling timelike observer to reach the singularity is infinite.

It seems that the existence of a curvature singularity in a vacuum, static, and spherically symmetric spacetime is still unavoidable in the mimetic Born-Infeld gravity, even though in some parameter space the original spacelike singularity in a black hole can be altered to a lightlike singularity. It would be interesting to include the angular momentum into the system and see how a Kerr black hole geometry would be changed in this theory. We leave this interesting issue for a coming work.
\acknowledgments

The work of MBL is supported by the Basque Foundation of Science Ikerbasque. She also wishes to acknowledge the partial support from the Basque government Grant No.~IT956-16 (Spain) and FONDOS FEDER under grant FIS2014-57956-P (Spanish government). CYC and PC are supported by Taiwan National Science Council under Project No. NSC 97-2112-M-002-026-MY3 and by Leung Center for Cosmology and Particle Astrophysics, National Taiwan University. CYC would like to thank the Department of Theoretical Physics of the University of the Basque Country for kind hospitality while part of this work was done. This article is based upon work from COST Action (CA15117, CANTATA), supported by COST (European Cooperation in Science and Technology).

\end{document}